# Topological pseudogap in highly polarizable layered systems with 2D hole-like dispersion

S. V. Doronkina, A. E. Myasnikova, A. H. Dzhantemirov and A. V. Lutsenko

*Faculty of Physics, Southern Federal University, 5 Zorge str., 344090, Rostov-on-Don, Russia*

Pseudogap in hole-doped cuprate superconductors is acknowledged as a possible key to understanding their ground normal state, however, existing pseudogap models have essential inconsistencies with experiments. In our approach pseudogap emerges due to impact of autolocalized carriers on stationary states of delocalized ones and topology of the 2D hole-like dispersion. Autolocalized carriers create potential which transforms Bloch quasiparticles into distributed wave packets (DWPs) with different momentums in areas with different potential. Topology of hole-like constant energy curves in 2D-conducting cuprates forbids DWPs with average momentums near antinode. Manifestation of permitted DWPs in ARPES spectra demonstrates all known pseudogap features including midpoint shift, giant broadening and Fermi momentum misalignment. Calculated doping dependence of the pseudogap width and onset temperature agrees with experiments. The obtained ground normal state of the hole-doped cuprates, in which all the doped holes are autolocalized until high doping is reached and near-antinodal electron DWPs are absent, explains Fermi surface reconstruction from small electron pocket to large hole-like Fermi surface observed in quantum oscillation measurements. Our results open a possibility of creating systems with artificial pseudogap and switchable density of states on the base of highly polarizable layered structures with 2D conductivity and hole-like dispersion.

**Keywords**: pseudogap, cuprates, charge ordering, 2D electron systems, Fermi surface topology, ARPES spectra

## 1. INTRODUCTION

Nature of pseudogap (PG) observed in hole-doped cuprate superconductors is seemingly as enigmatic as superconductivity in them [1-3]. The term PG is ordinarily used to denote absence of the carriers with momentums near antinodes, i.e. in the vicinity of points $(0,\pm\pi)$, $(\pm\pi,0)$ in the First Brillouin zone (FBZ), in some interval of energies near Fermi surface (FS). PG is observed at temperatures up to $T^*$ which is higher than the temperature of superconducting transition, except overdoped systems. The most detailed information about PG is provided by ARPES and STM methods [4-14], and recent ARPES data [6-8] have brought new constraints on theoretical models of the PG mechanism. Initially PG was related with scattering of carriers resulting in formation of charge/spin density wave (CDW/SDW) with the wave vector of the 2D FS nesting. However, these models turned out to be at variance with recent ARPES data [6,7] demonstrating particle-hole symmetry breaking and giant broadening of features that do not arise in conventional CDW picture [15,16]. Nevertheless, already these first models pointed out importance of 2D character and topology of the carrier dispersion in cuprates for the PG emergence.

The alternative Amperian pairing, or pair density wave (PDW), mechanism of the PG formation [15,16] yields particle-hole asymmetry but has another discrepancy with experiments on cuprates. In this approach pairing of near-antinodal carriers with large binding energy occurs, and it is not clear, why it does not result in superconductivity [15,16]. Besides, giant broadening of features in near-antinodal ARPES spectra [6,7] casts doubt on using states with certain momentums to describe near-antinodal carriers. In addition, recent experimental results on the role of FS topology in PG formation [17,3] and on absence of hole FS and, hence, delocalized holes, until rather high doping level is reached [18,3], indicate necessity to reconsider the PG problem in complex with the idea about ground normal state of cuprates.

Earlier giant broadening of nodal ARPES features was explained as an effect of strong electron-phonon interaction (EPI) [19-23], however, possible impact of EPI on antinodal carrier states and antinodal ARPES spectra has not been studied yet. Softening of optical phonons with the wave vectors near the charge ordering (CO) one [24,25] as well as wide bands in ARPES and optical spectra [19-23,26,27] are ordinarily considered as fingerprints of strong EPI in cuprates. Recently experimental evidence of strong EPI in cuprates was obtained with inelastic tunneling spectroscopy [28] and scanning noise spectroscopy [29]. Discovery of high-temperature superconductivity (HTSC) at the FeSe/SrTiO$_3$ interface [30-32] highlights the important role of EPI [33] in different HTSC systems. Therefore below we take EPI into account (on top of the electron correlations which determine the carrier dispersion in cuprates) and develop an approach in which PG emerges neither due to scattering of delocalized near-antinodal carriers as in CDW/SDW picture, nor due to their pairing as in the PDW approach.

Electronic correlations as well as strong short-range (Holstein) EPI are ordinarily considered in the node representation so that they are readily combined in one model [20,21,34,35]. Their concurrent consideration is necessary as the binding energy of the small polaron emerging at such EPI is of the order of the carrier bandwidth determined among others by strong electron correlations (in contrast with the large (bi)polarons case occurring at strong long-range EPI [36,37]). However, small polarons and bipolarons have very low mobility [36,37,26] and are limited in size by an elementary cell thus excluding coexistence with delocalized carriers, in distinct from large (bi)polarons. As a result, conductivity of systems with small (bi)polarons is orders of magnitude lower than that observed in cuprates. Besides, strongly ionic lattice of cuprates favors long-range (Fröhlich) EPI [37,26]. Therefore, here we consider systems with strong Fröhlich EPI. Binding energy of large polarons and bipolarons emerging in such systems is much smaller than the carrier bandwidth, so that it is reasonable to take such EPI into account after the electron correlations which determine the carrier dispersion in cuprates [38].

The PG problem is inseparably connected with the problem of ground normal state of hole-doped cuprates [1-4,17,18]. The ground normal state of a system with strong Fröhlich EPI and high carrier density obtained with generalized variation method represents two-component liquid where compressible large bipolarons (which are much larger than the unit cell [39,40]) form CO and coexist with delocalized carriers [41]. This approach has allowed calculating the CO wave vector (as it is related with the bipolaron size) and CO onset temperature [41] as functions of doping being in consent with those observed in cuprates [42-44]. The two-component model has also enabled calculating the high-energy anomaly ("vertical dispersion" and extremely broad features) in ARPES spectra [23] being in quantitative agreement with experiments on cuprates [45-49], the "vertical dispersion" there results from dividing the momentum space between autolocalized and delocalized carriers. Earlier various two-component models of cuprates' electronic structure were suggested to explain temperature behavior of the Hall constant [50,51].

Here we consider stationary states of delocalized carriers in presence of potential created by autolocalized ones (also called CO potential) and demonstrate the PG emergence. The developed two-component model also advances understanding the ground normal state of the hole-doped cuprates, in particular, the problem of FS reconstruction can be addressed in it. Recent quantum oscillation experiments revealed famous electron pocket and no sign of hole pockets at doping $p$ lower than some certain value $p^*$ and large hole-like FS at $p>p^*$ [18,3]. The electron pocket was successfully explained as a PG consequence [18], however, absence of hole FS at $p<p^*$ and its sudden appearance at $p=p^*$ remained enigmatic. The proposed approach sheds light on this puzzle: at low doping all the holes are autolocalized, so that delocalized holes are absent in the system until the doping exceeds $p^*$. The approach also addresses the long-time discussion of relation between CO and PG [1-3]. The suggested relation is in harmony with observed sequence of CO and PG onset temperatures $T_{CO} < T^*$ [3], since for CO manifestation a larger amount of survived (against thermal decay) bipolarons is needed than for PG one.

## 2. MODEL AND METHODS
### 2.1. Potential of the charge ordering

As was revealed experimentally [52] charge density in the CO phase of cuprates can be described as:

$$\rho = \rho_0\big(cos(K_{CO\,x}x) + cos(K_{CO\,y}y)\big), \tag{1}$$

where $\boldsymbol{K_{CO}}$ is the CO wave vector, $K_{CO\,x}= K_{CO\,y}=2\pi/C$, $C$ is the CO period along $x$ or $y$ axes coinciding in the present approach with the bipolaron size in these directions [41]. The amplitude $\rho_0$ is determined from equality of integral of (1) over the bipolaron volume to the bipolaron charge $2e/\varepsilon_0$ [53], where $\varepsilon_0$ is static dielectric constant. In the hole-doping case the bipolaron section by the conducting plane can be approximated with 90º rhombus with the diagonals $C$ along $x$ and $y$ axes for both electron and hole bipolarons [41]. In the direction perpendicular to the conducting plane the charge distribution is supposed uniform inside the layer of the height $h_1+h_2$ ($h_1$ and $h_2$ are determined by distances between the conducting layers, they may be different, especially in the case of several conducting layers in the unit cell). One can easily find amplitude of the carrier potential energy in the CO potential as its value in the points of minimum and maximum of the potential created by the charge distribution (1) (written in the cylindrical coordinates):

$$U_0 = e \sum_{j=0}^{n-1} \int_{-h_{1,j}}^{h_{2,j}} \int_0^{r_0} \int_0^{2\pi} \frac{(-1)^j 2\pi \rho r d\phi dr dz}{\sqrt{r^2+(z_{0j}+z)^2}}, \tag{2}$$

where n is doubled number of CuO planes in the unit cell (as calculation shows only taking nearest neighbor CuO planes into account influences the potential amplitude), $z_{0j}$ is distance from zeroth plane where the carrier is to $j$-th plane, $r_0$ is chosen sufficiently large so that its value does not influence $U_0$, and energetically profitable vertical alternation of the charge sign in adjacent layers is used. The $U_0$ value calculated according to (1,2) is obviously depending on the $K_{CO}$ value and, thus, on the doping $p$, as $K_{CO}$ in cuprates depends on doping [44], in all the cases $U_0$ is much smaller than the carrier bandwidth in cuprates.

Potential created by bipolarons (briefly denoted below as CO potential) for a carrier propagating in the conducting plane can be considered as a periodic function albeit with short coherence length. We consider below two such CO potentials (we will use term potential in the sense of potential energy unless the sign is important). The first is proportional to a harmonic function similar to (1) with phase shifted for gradual increase of the potential upon entering area with CO at $x=0$:

$$U = U_0\big(sin(K_{CO\,x}x) + sin(K_{CO\,y}y)\big)/2 = U_0(sin(2\pi x/C) + sin(2\pi y/C))/2, \tag{3}$$

To demonstrate that the shape of the potential does not affect the PG we calculate it also for the following CO potential

$$U = U_0[(sign(sin(K_{CO\,x}x))|sin(K_{CO\,x}x)|^{\frac{1}{4}} + sign(sin(K_{CO\,y}y))|(sin(K_{CO\,y}y)|^{\frac{1}{4}}]/2. \tag{4}$$

Fig. 1 represents both shapes of the CO potential (3) and (4). We have found that the periodicity of CO is not necessary for the effect: we have obtained the same PG features using CO potential with frustrated periodicity, the coherence length used was the same as one observed in cuprates.

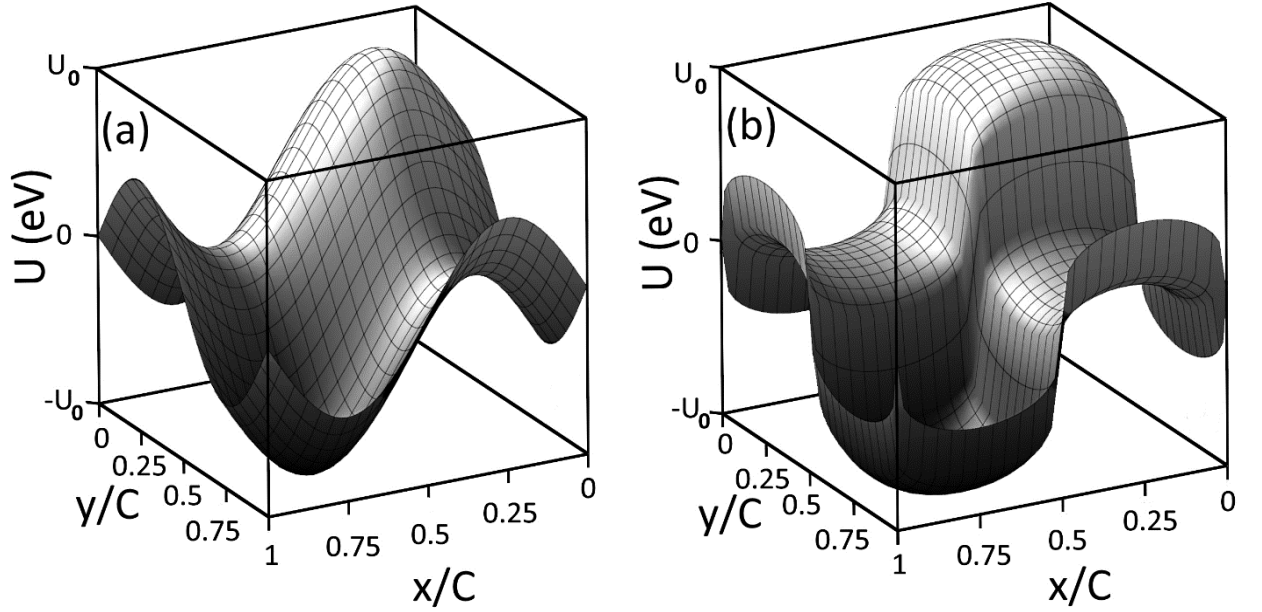

FIG. 1. (a) and (b) CO potential set by Eq. (3) and Eq. (4), respectively.

To unveil how the potential created by autolocalized carriers transforms the states of delocalized ones one should solve Schrödinger equation with the additional potential generated by bipolarons. As near-antinodal carriers have one of two projections of the wave vector onto the conduction plane much larger than the other, we consider below the states with $k_x \gg k_y$. Let us first neglect change of the CO potential in y direction to begin from the simpler and more transparent consideration and then study the case of CO potential depending on both $x$ and $y$. We divide the conduction plane into $N = N_p N_l + 1$ equal stripes parallel to y axis, where $N_p$ is the number of the CO periods in the structure and $N_l$ is the number of stripes in the CO period, and average the bipolaron potential (3) in each (j-th) layer over the CO period along y axis:

$$\langle U_j \rangle = \frac{1}{2R_{bip}} \int_0^{2R_{bip}} U(x_j, y)\, dy = U_0 \sin(K_{COx} x_j), \qquad (5)$$

where $U_0$ is now the amplitude of the averaged potential. Similar procedure applied to potential (4) yields

$$\langle U_j \rangle = U_0 sign(\sin(K_{COx} x_j)) |\sin(K_{CO\,x} x_j)|^{1/4} \qquad (6)$$

The stripe width $s=C/N_l$ is chosen such that the potential can be considered constant in the limits of each stripe. It can be not small: as is seen from Fig.1(b) it can coincide with the bipolaron radius.

**2.2. Searching stationary states of delocalized carriers in CO potential**

To consider transformation of Bloch carrier state with the 2D quasi-momentum **k₀** in presence of additional CO potential let us search solution $\Psi_j$ of Schrödinger equation for stationary states of a carrier in the crystal periodic potential and additional potential $U_j$ constant in the j-th layer ($U_j$ is small in comparison with the bandwidth of the lower Hubbard band and bandgap)

$$\left[-\frac{\hbar^2 \nabla^2}{2m} + V + U_j\right]\Psi_j = E(\mathbf{k}_0)\Psi_j$$

corresponding to independent of $j$ energy $E$ depending on $\mathbf{k_0}$. It is natural to search $\Psi_j$ as expansion in terms of Bloch wave functions of the lower Hubbard band: $\Psi_j = \sum_\mathbf{k} C_{\mathbf{k},j}\psi_\mathbf{k}$. Then using Schrödinger equation for Bloch wave function, one obtains

$$\sum_\mathbf{k}[\varepsilon(\mathbf{k}) + U_j - E(\mathbf{k}_0)]C_{\mathbf{k},j}\psi_\mathbf{k} = 0,$$

where $\varepsilon(\mathbf{k})$ is Bloch electron dispersion. As the boundaries between the layers with constant potential are parallel to $y$ axis, the $y$-projection of the wave vector $k_y$ is conserved on each boundary and thus is constant:

$$k_{j,y} = k_j \sin\alpha_j = k_0 \sin\alpha_0, \tag{7}$$

where $k_j$, $\alpha_j$ и $k_0$, $\alpha_0$ are the modulus of the wave vector and the angle between it and $x$ axis (shown in Fig.2(a)) in the $j$-th layer with the CO potential $U_j$ and in a layer with zero CO potential, respectively. Since the symmetry of the momentum space coincides with the symmetry of the coordinate space, in the tetragonal case under study the coefficients $C_{\mathbf{k},j}$ differ from zero only at $\mathbf{k}_j = (k_{j,x}, k_{j,y})$ and at $\mathbf{k}_j' = (-k_{j,x}, k_{j,y})$ shown in Fig.2(a), where $k_{j,y}$ is determined by (7) and $k_{j,x}$ satisfies the equation

$$\varepsilon(\mathbf{k}_j) = \varepsilon(\mathbf{k}_j') = E(\mathbf{k}_0) - U_j. \tag{8}$$

Therefore, $\Psi_j = C_{\mathbf{k}_j}\psi_{\mathbf{k}_j} + C_{\mathbf{k}_j'}\psi_{\mathbf{k}_j'}$, where coefficients $C_{\mathbf{k}_j}$, $C_{\mathbf{k}_j'}$ are determined by standard boundary conditions relating carrier's wave functions in adjacent layers:

$$\begin{cases}\Psi_{j-1}(x_j, y_j) = \Psi_j(x_j, y_j) \\ \left.\frac{\partial \Psi_{j-1}}{\partial x}\right|_{x_j,y_j} = \left.\frac{\partial \Psi_j}{\partial x}\right|_{x_j,y_j}\end{cases}, \tag{9}$$

$x_j = x_{j-1} + s$, $y_j = y_{j-1} + s * tg\alpha_j$, $tg\alpha_j = k_{j,y}/k_{j,x}$. As is seen from (8), in layers with zero CO potential $E(\mathbf{k}_0) = \varepsilon(\mathbf{k}_0)$, so that $\mathbf{k}_0$ and $\mathbf{k}_0'$ are quasi-momentums of the new QP in such layers.

Thus, the obtained carrier stationary state in additional CO potential has different quasi-momentums in areas with different CO potential: although $y$-projection of the quasi-momentum is conserved at the boundaries of the layers with constant potential (since $U_j$ (5,6) does not depend on $y$ coordinate), $x$-projection of the quasi-momentum changes its value according to Eq. (8), as is illustrated by Fig. 2(a). It can be noted that distribution of the QP state over wave vectors resembles a wave packet. However, in this distributed wave packet (DWP) the components with various wave vectors are present in layers of the coordinate space with various CO potential.

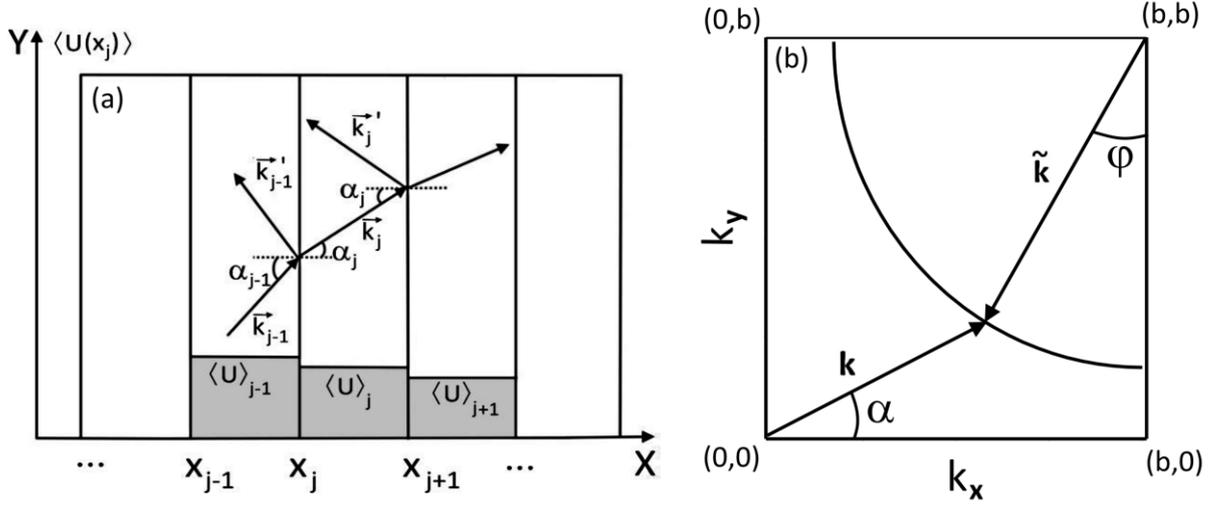

FIG. 2. (a) Wave vectors of the incident and reflected electron waves in several layers with constant (in the layer) potential; (b) relation of the wave vector **k** and angle α (between **k** and x axis) and the vector $\tilde{k}$ and angle φ; the arc of circumference is the constant energy $E$ surface, characteristic of cuprates' dispersion, $b=\pi/a$.

Of course, approximation of the CO potential with Eqs. (5,6) is more suitable near antinode where *y* projection of the wave vector is much shorter than its *x*-projection so that change of the potential along *y* axis during the carrier propagation may be initially neglected. Such simplified model allows demonstrating more apparently the peculiarities of new QPs in a system with cuprates-like dispersion. Then we consider the case of the CO potential depending on both *x* and *y* coordinates.

Modeling carrier propagation in a CO potential depending on both coordinates is carried out in a way similar to one described above. The square region of the conducting plane with the side $CN_p$ is divided into square cells with the sides $s = C/N_l$. The potential in each cell is considered constant calculated in the case of CO potential (3) as follows:

$$U_j = \frac{U_0}{2} * \left[\sin\left(\frac{2\pi X_j}{C}\right) + \sin\left(\frac{2\pi Y_j}{C}\right)\right], \tag{10}$$

where $X_j$, $Y_j$ are the coordinates of the bottom left angle of *j*-th cell. Comparison of the angle $\alpha_j$ with the limiting angle $\alpha_l$,

$$\alpha_l(j) = arctg\left[(Y_j + y_j - s)/(X_j + x_j - s)\right] \tag{11}$$

determines whether reflection and refraction of the Bloch waves occur at a cell boundary parallel to *x*-axis or *y*-axis. If the refraction and reflection occur on the boundary parallel to *x* axis then in distinct from Eqs.(7,8) the *x*-projection of the quasi-momentum is conserved: $k_{j,x}=k_{j+1,x}$, the *y*-projection of the wave vector changes its value according to (8), and the boundary conditions (9) include continuity of $d\Psi/dy$ on the boundary between layers. Therefore, the Eq.(7) should be replaced in this general case with the following condition:

$$k_{j,y}=k_{j+1,y},\ \alpha_j < \alpha_l(j), \quad k_{j,x}=k_{j+1,x},\ \alpha_j > \alpha_l(j) \tag{12}$$

Thus, stationary state of a carrier in additional CO potential created by autolocalized carriers is DWP with different momentums in areas with different potential: during propagation in additional CO potential the carrier quasi-momentum should satisfy the system of Eqs. (8) and (12). To set a DWP state in the considered 2D system one can determine either the wave vector projections ($k_{x0}$, $k_{y0}$) in a layer with zero CO potential or one of the values $\tilde{k}$, $E$, or $k_{y0}$ together with the angle $\varphi_0$ (often used in cuprates and shown in Fig. 2(b)) in a layer with zero CO potential.

### 2.3. Modeling cuprates dispersion in the vicinity of FS and antinodes

As distribution of carrier stationary states in CO potential over wave vectors is determined by Bloch carrier dispersion $\varepsilon(\mathbf{k})$ let us find suitable expression modeling near-antinode dispersion in the lower Hubbard band of cuprates. The dispersion can be taken from that of ARPES spectra of cuprates since as was shown earlier [20,21] in systems with strong EPI it follows "bare" (i.e., in absence of EPI) carrier dispersion. Experimental ARPES data on cuprates show that width of the lower Hubbard band is about 0.5 - 0.6 eV [4,45] and the Fermi surface (FS) at different doping levels *p* typical of cuprates can be approximated by four 90- (or slightly lower) degree arcs [4,46]. To obtain 90-degree arcs we put their center (for the first quadrant of the FBZ) in ($b$,$b$) point, where $b = \pi/a$, $a$ is the lattice constant, as is shown in Fig. 2(b). From the condition of equality of the filled and unfilled areas in the FBZ at zero doping the FS radius at zero doping is $\tilde{k}_0 = b\sqrt{2/\pi}$, at higher doping the FS radius $\tilde{k} > \tilde{k}_0$. Tildes mark that these vectors of the momentum space start in the point (b,b). Wave vectors $\mathbf{k}=(k_x,k_y)$ corresponding to points on a constant energy surface are simply related with its radius $\tilde{k}$: $\tilde{k}^2 = (k_x - b)^2 + (k_y - b)^2$, as is illustrated by Fig. 2(b).

Modeling the cuprates' dispersion, we tried to reflect its flatness near FS in the vicinity of the antinode [4,38,54]. We study several dispersion laws of the form (13) with different *d* values from 1 to 2 that satisfy the conditions described above and are in rather good accordance with experimental dispersion in cuprates near the antinode under hole doping *p*<0.26:

$$\varepsilon = 0.5 - c * (\tilde{k} - \tilde{k}_0)^d, \tilde{k} > \tilde{k}_0, \quad \varepsilon = 0.5 + c' * (\tilde{k}_0 - \tilde{k})^{d'}, \tilde{k} < \tilde{k}_0, \tag{13}$$

As will be seen below the dispersion slightly above the zero-doping FS is also necessary for determining QPs in additional CO potential, therefore in (13) we continue the dispersion above the zero-doping FS preserving its flatness near the antinode. Fig. 3(a,b) presents dispersion (13) for *d*=1.2 and 1.5, respectively. It should be noted that we do not complicate dispersion (13) to achieve similarity with cuprates' dispersion in nodal direction as it is not used in the further study.

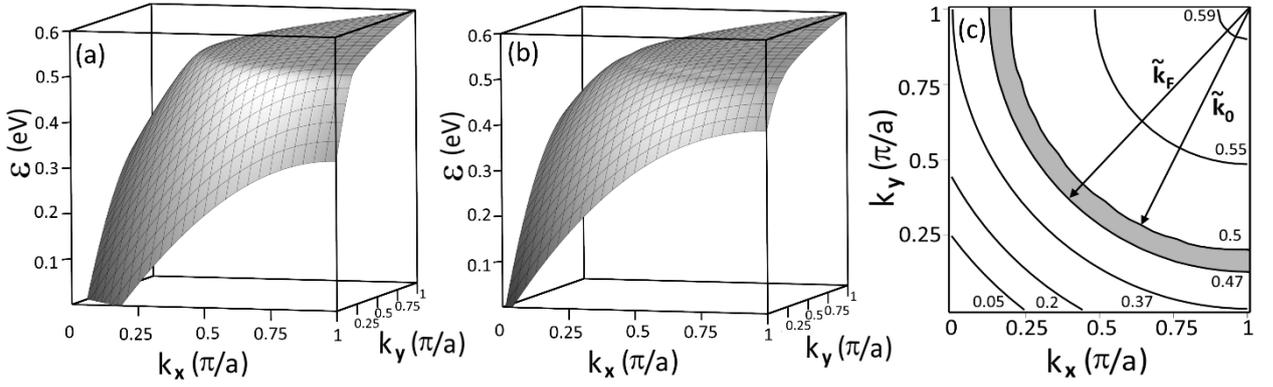

FIG. 3. (a) and (b) Electron dispersion (13) with $d=1.2$ $c=1.5$, $d'=0.65$, $c'=0.13$ and $d=1.5$, $c=1.5$, $d'=0.65$, $c'=0.13$, respectively; (c) Fermi energy determination from the area between the constant-energy curves for the case of doping $p=0.2$ for dispersion (13) at $d=1.5$, $c=2$, $d'=0.65$, $c'=0.13$, Fermi energy changes from 0.499 eV at $p=0.01$ to 0.458 eV at $p=0.25$. Curves are the constant energy $\varepsilon$ surfaces, numbers near the curves indicate $\varepsilon$ value for each curve. Grey region depicts the area $S$.

To calculate the PG width, one needs Fermi energy as function of doping. In the quasi-2D system (with in-plane conductivity) considered here it is easily deduced from the equation relating the area $S$ in k-space between the FSs corresponding to zero and non-zero doping (shown with grey in Fig. 3(a,b)) and the surface density of doped holes $n_h$: $S = (2\pi)^2 n_h/2$, where $n_h = p/a^2$, $p$ is the doping value, and 2 is due to spins. In the case of dispersion (13) $S$ is the area between two arcs, which are cross sections of the surface $\varepsilon(\mathbf{k})$ by the planes $\varepsilon(\mathbf{k}) = \varepsilon_F$ and $\varepsilon(\mathbf{k}) = \varepsilon(\tilde{\mathbf{k}}_0)$. $S$ is the area of a ring of the width $\Delta \tilde{k} = \tilde{k}_F - \tilde{k}_0$, where $\tilde{k}_F = \sqrt{2\pi n_h + \tilde{k}_0^2}$, its quarter is demonstrated by Fig. 3(c). Thus, for dispersion (13) Fermi energy as function of doping has the form $\varepsilon_F(p) = 0.5 - c(\tilde{k}_F(p) - \tilde{k}_0)^d$.

## 3. RESULTS AND DISCUSSION

### 3.1. Vanishing near-antinodal DWPs at cuprates-like dispersion

Let us consider limitation on near-antinodal DWPs in a system with cuprates-like dispersion. First we discuss a simpler case taking the inhomogeneity along one axis into account that is reasonable approximation for DWPs with near-antinodal average momentum (then we consider the general case of the CO potential depending on both, $x$ and $y$, coordinates). As was shown above, during propagation in additional CO potential depending on $x$ coordinate the carrier quasi-momentum should satisfy the system of Eqs. (7) and (8). Substituting dispersion (13) into Eq. (8) one obtains

$$E - U_j = 0.5 - c * \left( \sqrt{(k_{j,x} - b)^2 + (k_{j,y} - b)^2} - \tilde{k}_0 \right)^d, E - U_j < 0.5,$$

$$E - U_j = 0.5 + c' * \left( \tilde{k}_0 - \sqrt{(k_{j,x} - b)^2 + (k_{j,y} - b)^2} \right)^{d'}, E - U_j > 0.5. \quad (14)$$

Thus, in the case of carrier dispersion (13) the QP wave vector should satisfy Eqs. (7,14).

Fig. 4 demonstrates trajectories along which quasi-momentum of near-antinodal QPs changes at their propagation in additional CO potential like (5) or (6). Geometrical analysis of compatibility of Eqs. (7,14) illustrated by Fig. 4 seems the most apparent. Let us consider stationary states with the total energy $E_1$ located on the arc shown in Fig. 4(a). At the QP propagation the potential energy $U_j$ takes different values, so that the left-hand side of Eq. (14) changes from $\varepsilon=E_1-U_0$ to $\varepsilon=E_1+U_0$ (also shown in Fig.4(a)). Concurrently $k_x$ value in the right-hand side should change at fixed $k_y$ value equal to y-projection of the wave vector in zero-potential layers (ZPL) according to (7). As was mentioned above, the QP states in additional CO potential can be specified not only by the momentum projections $(k_{x0}, k_{y0})$ in zero-potential layers (ZPLs) but also by the energy $E$ (which determines the arc radius in the momentum space) and angle $\varphi_0$ in the ZPLs: $(E, \varphi_0)$. Let us compare trajectories in the momentum space of two near-antinodal QPs with the same energy $E_1$ but two different wave vectors in ZPLs specified by angles $\varphi_1$ and $\varphi_2$ in Fig. 4(a). The first trajectory (line 1 in Fig. 4(a)) corresponding to the larger angle $\varphi_1$ reaches both the minimal kinetic energy surface $\varepsilon=E_1-U_0$ (this occurs in the layers with the maximal potential energy, point A in Fig. 4(b)) as well as the maximal kinetic energy surface $\varepsilon=E_1+U_0$ (in the layers with the minimal potential energy, point C in Fig. 4(b)). The second trajectory (line 2 in Fig. 4(a)) corresponding to smaller angle $\varphi_2<\varphi_1$ reaches the former surface but does not reach the latter.

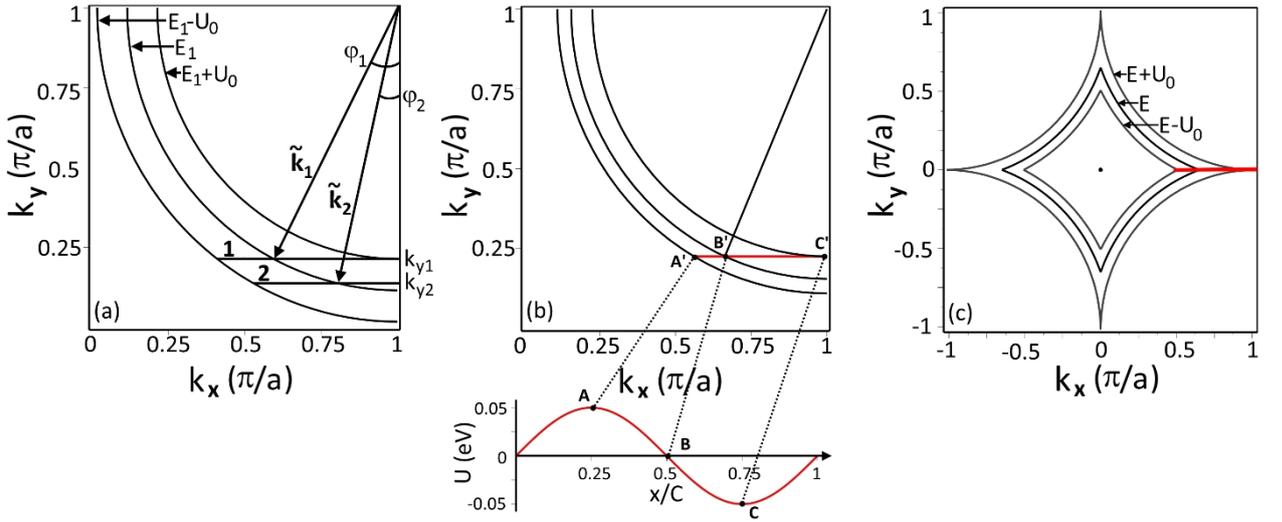

FIG. 4. (a) Lines 1,2 depict trajectories of the carrier momentum at propagation in additional CO potential. While trajectory 1 reaches both maximum $\varepsilon=E_1+U_0$ and minimum $\varepsilon=E_1-U_0$ kinetic energy surfaces and, thus, correspond to real QPs, trajectory 2 does not reach the surface $\varepsilon=E_1+U_0$, therefore carrier state $(E_1, \varphi_2)$ is not a real QP. (b) Structure of the new QP state (DWP): correspondence between points of trajectory in the momentum space and potential energy layers where the carrier has such momentum value; (c) Illustration of PG disappearance when the constant energy surface $\varepsilon=E+U_0$ becomes electron-like in agreement with experiment [17]: momentum trajectory (red line) of the antinodal carrier with the energy E reaches both maximum $\varepsilon=E+U_0$ and minimum $\varepsilon=E-U_0$ kinetic energy surfaces, for $d=1.5$, $c=1.5$, $E=0.35$ eV.

Hence, at $\varphi \geq \varphi_1$ real roots of the system of Eqs. (7,14) exist in each layer of the CO potential. Therefore, the states with the energy $E_1$ and $\varphi \geq \varphi_1$ are real QPs. For the case shown in Fig.4(a,b) at $\varphi_0<\varphi_1$ and $E=E_1$ there are no real roots of the system (7,14) in some layers with negative carrier potential energy. Damping of such states caused by imaginary part of the wave vector results in their absence among new QPs. Thus, negative half-wave of the additional potential energy caused by attraction of electrons to hole bipolarons together with topology of the constant energy surfaces characteristic of cuprates result in vanishing of the near-antinodal QPs. It should be noted that, due to partial reflection of electron waves, besides the trajectory in the first quadrant of the FBZ shown in Fig. 4, complete momentum trajectory comprises the trajectory reflected in plane which is perpendicular to $x$ axis and contains (0,0) point. Below we will consider how the permitted QPs display themselves in experiments. Fig. 4(c) demonstrates disappearance of the described above limitation on the near-antinodal QPs when the constant energy surfaces that bound the momentum trajectory becomes electron-like as it was observed experimentally by N. Doiron-Leyraud et al. [17].

### 3.2. Manifestation of DWPs in antinodal ARPES spectra

#### 3.2.1. Shifting spectral weight to higher binding energy by the CO potential amplitude

Let us consider how the revealed transformation of stationary states of delocalized carriers in presence of autolocalized ones and limitations on new QPs in systems with cuprates-like dispersion display themselves in ARPES spectra. Obviously, an electron with given total energy $E$ (which is equal to its kinetic energy $\varepsilon$ only in layers with zero CO potential) can escape due to photoabsorption from areas with different potential energy $U_j$ and, as a result, can have different values of the wave vector (and $\varphi$) satisfying the Eqs. (7, 14). This is illustrated in Fig. 4(b) depicting a trajectory of the QP wave vector and the potential energy corresponding to some trajectory points.

The QP with the energy $E=E_1$ and $\varphi_0=\varphi_2$ in Fig. 4(a) does not exist, as was discussed above, and, consequently, does not display itself in ARPES spectrum. The QP with the energy $E=E_1$ and $\varphi_0=\varphi_1$ in Fig. 4(a) (having $y$-projection of the wave vector $k_{y1}$) exists. Its trajectory in the momentum space achieves the curve of constant kinetic energy $\varepsilon=E_1+U_0$ at the FBZ boundary, i.e. at **k**=$(b,k_{y1})$:

$\varepsilon(b,k_{y1})=E_1+U_0$

Therefore, in the antinode the carrier total energy $E_1$ is $U_0$ lower than it is dictated by the dispersion law for its momentum: $E_1=\varepsilon(b,k_{y1})-U_0$. If this carrier being in the layer with the minimum potential energy (point C in Fig. 4(b)) absorbs a photon then it displays itself in the ARPES spectrum with the momentum **k**=$(b,k_{y1})$ and binding energy $E_1-E_F = \varepsilon(b,k_{y1})-U_0-E_F$.

Thus, photoelectrons with the antinodal wave vectors and corresponding to them (according to the dispersion law) binding energy $\varepsilon(\mathbf{k})-E_F$, ordinarily observed in normal metals and in cuprates at $T>T^*$, are absent in the considered system. Instead spectral weight near the FBZ boundary **k**=$(b,k_y)$ appears at higher (by the potential energy amplitude $U_0$) binding energy $\varepsilon(\mathbf{k})-U_0-E_F$. Such shift of the spectral weight to higher binding energies by approximately 0.05 eV in comparison with the spectrum taken at temperature $T>T^*$ (when CO is absent) is indeed observed in the near-antinodal ARPES spectra [6,7]. The value of the midpoint shift in the binding energy at $k_y=k_{Fy}$, where $k_{Fy}$ is Fermi crossing at $T>T^*$, is ordinarily considered as PG width, so that in the present approach it coincides with $U_0$. Shift of the dispersion in the antinodal region by $U_0$ value down in the binding energy results in appearance of the spectral weight at momentums $k_y>k_{Fy}$ and binding energies from $\varepsilon(\mathbf{k})-U_0-E_F$ up to $\varepsilon(\mathbf{k})-E_F$. Such spectral weight was observed in experiments [6,7] where it was denoted particle-hole symmetry breaking, or $k_F$ misalignment [15]. Giant broadening of the spectral weight and decrease of intensity with decreasing binding energy at $k_y>k_{Fy}$ characteristic of experimental spectra [6,7] also emerge in the considered system if dependence of the CO potential on both coordinates *x* and *y* is taken into account as it is demonstrated below.

### *3.2.2. Angular dependence of the PG*

Let us consider angular dependence of the PG in the suggested approach and show that it is also in agreement with that observed in cuprates [4, 6-8]. As we have seen above at the antinode (at $\varphi=0$) the PG width is $U_0$. From the other hand, as is seen from Fig. 4(a), there exists the minimal value of $\varphi_0$ corresponding to real QP with given energy. It corresponds to closing the PG and can be denoted as $\varphi_{clos}$. For example, in Fig. 4(a) $\varphi_{clos}=\varphi_1$ provided $E_F=E_1$. Obviously, motion along the FS towards the increase of $\varphi$ results in decreasing PG from $U_0$ at $\varphi=0$ to zero at $\varphi_{clos}$. The value of $\varphi_{clos}$ is related with the dispersion parameters. Geometrical demonstration of this relation is illustrated by Fig. 5(a). For instance, let us find an example of the dispersion corresponding to $\varphi_{clos}=30°$ like it is observed in experiments on cuprates. To do this one should solve an Eq.

$$\tilde{k}(E)\cos 30° = \tilde{k}'(E+U_0), \qquad (15)$$

where $\tilde{k}(\varepsilon)$ and $\tilde{k}'(\varepsilon)$ are determined by the dispersion (13) and are shown in Fig. 5(a). For example, for $p=0.15$ at $d=1.2$, $c=1.5$ and $c'=0.1$ the solution of Eq. (15) yields $d'=0.8$. Figs. 5(b,c) demonstrate the Fermi-surface angles corresponding to PG closing for systems with various dispersion parameters and doping levels *p*.

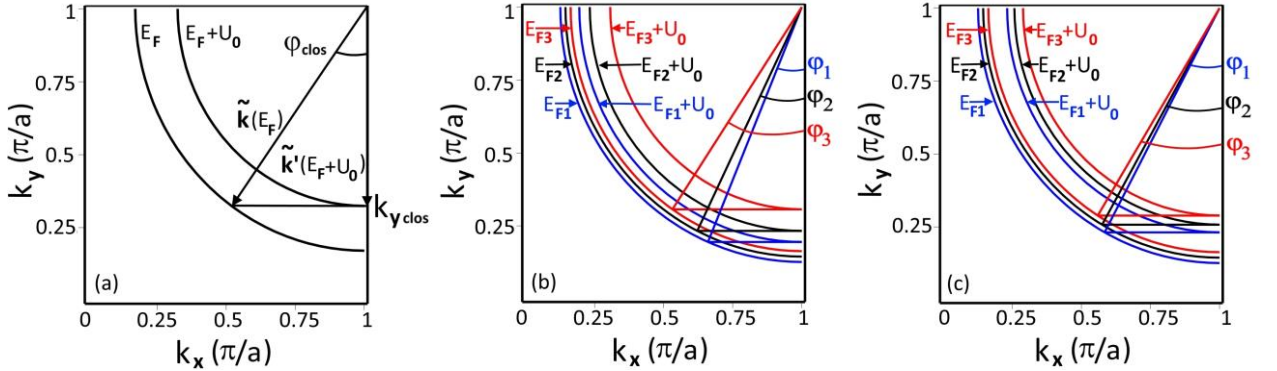

FIG. 5. (a) The method to calculate the parameter *d'* in dispersion (13) corresponding to PG closing at given Fermi-surface angle $\varphi_{clos}$; (b, c) variation of the closing angle with the doping *p* and dispersion parameters: in panel (b) $d=1.2$, $c=1.5$, $d'=0.65$, $c'=0.13$, $E_{F1}(p=0.2)=0.447$, $\varphi_1=23°$, $E_{F2}(p=0.15)=0.461$, $\varphi_2=26°$, $E_{F3}(p=0.1)=0.476$, $\varphi_3=33.6°$; in panel (c) $d=1.5$, $c=1.5$, $d'=0.4$, $c'=0.12$, $E_{F1}(p=0.2)=0.477$, $\varphi_1=28.5°$, $E_{F2}(p=0.15)=0.485$, $\varphi_2=30°$, $E_{F3}(p=0.1)=0.492$, $\varphi_3=32°$.

### *3.2.3. Broadening ARPES features at CO potential depending on both x and y coordinates*

Let us now study a more general model without averaging the CO potentials (3,4) over *y* coordinate. We still consider near-antinodal carriers but now we take into account that propagating Bloch waves face also boundaries parallel to *x* axis (in the present approach all boundaries are chosen to be parallel to the coordinate axes). As a result, their trajectories in the momentum space cease to be horizontal lines and consist of horizontal and vertical small steps enclosed in quasi-rectangles exemplified by Fig. 6. During propagation of such DWP quasi-momentum changes according to the system of Eqs.(12,14).

As is seen from Fig. 6(c,f), trajectories of near-antinodal QPs in the momentum space achieve the surfaces of the highest and lowest kinetic energy approximately along the diagonal of the quasi-rectangle, the angle between the diagonal and the *x*-axis can be approximated with α. Thus, one can again obtain geometrical estimation of the dispersion parameters providing the PG closing at any given Fermi-surface angle $\varphi_0$. Besides $\varphi_0$, the equation will include $k_{x0}$ value shown in Fig. 6(c,f), although for rough estimation one can use dispersion parameters satisfying the Eq. (15) with $\varphi_0$ instead of 30°. Comparison of panels (a) and (d), (b) and (e), (c) and (f) in Fig. 6 shows that QP momentum-space trajectory does not depend on the form of CO potential, which is determined by Eq. (3) for panels (a)-(c) and by Eq. (4) for panels(d)-(f).

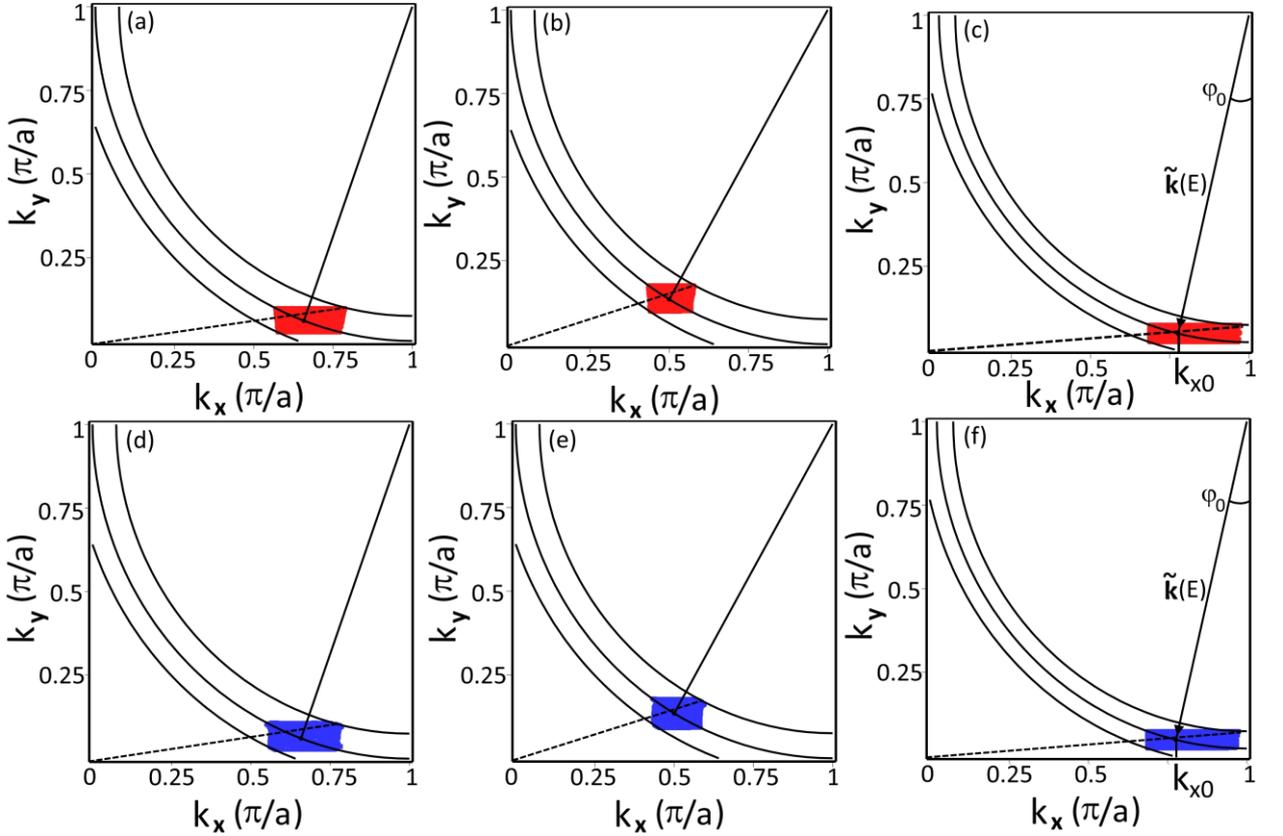

FIG. 6. Trajectories of the QP wave vector at propagation in CO potential (3) (a)-(c) and (4) (d)-(f) with the amplitude $U_0$=0.05 eV; $N_p$=100, $N_l$=32 (panels a-c), $N_p$=4000, $N_l$=256 (panels d-f). In panels (a) and (d) *d*=1.5, *c*=1.5, *E*=0.4 eV, $\varphi_0$=20°; in panels (b) and (e) *d*=1.5, *c*=1.5, *E*=0.4 eV, $\varphi_0$=30°; in panels (c) and (f) *d*=1.2, *c*=1.5, *E*=0.35 eV, $\varphi_0=\varphi_{clos}$=13.2°.

In the frames of the suggested approach with CO potential depending on both *x* and *y* coordinates giant broadening of the EDC maximums in ARPES spectra observed in experiments [6,7] emerges naturally as is illustrated by Fig. 7 (a,b). It represents momentum trajectories for two QPs, whose energies $E_1$ and $E_2$ differs by more than $U_0$. The trajectories intersect or touch each other, therefore photoelectrons with the momentums from the intersection of red and blue regions can have any energy from the interval [$E_1$, $E_2$] which is wider than $U_0$. Similar broadening is demonstrated by near-antinodal spectral weight in ARPES spectra of cuprates at $T<T^*$ [7,6]. It should be noted that complete trajectories of the QP momentum apart of those shown by Figs. 6,7 comprise also ones reflected in planes containing $k_x$ or $k_y$ axes and perpendicular to the ($k_x,k_y$) plane.

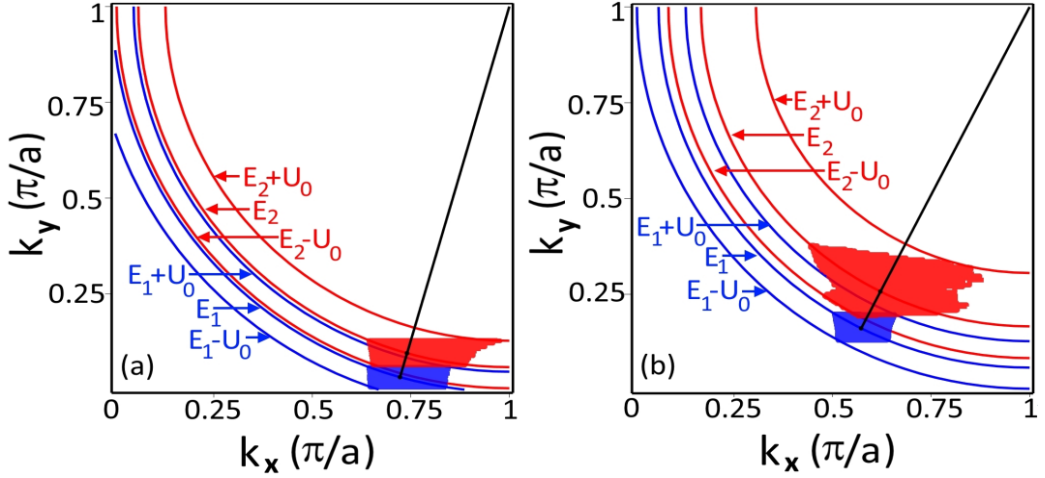

FIG. 7 (a),(b). Illustration of source of the ARPES features broadening. Blue and red areas are trajectories of the momentum of QPs with the same $\varphi_0$ but different energies $E_1$ and $E_2$, respectively, $E_2-E_1>U_0$. Photoelectron escaping from crystal with the wave vectors from the intersection of red and blue regions can have any energy from the interval [$E_1$, $E_2$] whose width is larger than $U_0$. For both panels $d=1.5$, $c=2$, $d'=0.65$, $c'=0.2$, $N_p=100$, $N_l=32$; in panel (a) $\varphi_0=16$, $E_1=0.36$ eV, $E_2=0.42$ eV; in panel (b) $\varphi_0=27$, $E_1=0.42$ eV, $E_2=0.49$ eV.

### 3.3. Calculated ARPES spectrum at CO potential depending on both x and y coordinates: PG, giant broadening, and features above $k_F$

To calculate antinodal ARPES spectrum in the frames of the present approach it is necessary first to obtain intersections of the QP momentum-space trajectories with the FBZ boundary. As Fig.7(a) demonstrates, in the antinode the quasi-rectangular QP momentum-space trajectory degenerates into a line resembling the 1D case, shown in Fig.4(a). If, however, instead of 90-degree arcs the constant energy curves according to the dispersion law are arcs leaning on a smaller angle (as it is observed in ARPES spectra of cuprates [48,55,56]) then trajectories remain quasi-rectangular up to the antinode at lower QP energies and degenerate into a line near antinode for QPs with energies closer to Fermi energy (due to much wider rings between curves $\varepsilon=E-U_0$ and $\varepsilon=E+U_0$ near $E=E_F$ caused by flatness of the cuprates-like dispersion near FS in the antinode). To obtain constant kinetic energy curves in the form of arcs with angle slightly smaller than 90° we substitute coordinates of the arcs' center $b$ with a slightly (by 13% here) larger $b'$ as is demonstrated by Fig. 8(a).

As the binding energy in ARPES spectrum is counted from the Fermi energy the second step is determining the FS position. Obviously, low-temperature (at $T<T^*$) FS and high-temperature (at $T>T^*$) FS do not coincide due to absence of QPs with average momentums near antinode at $T<T^*$ where bipolarons and, therefore, PG are present. To find the low-temperature FS corresponding to zero doping let us take into account that at $T<T^*$ some carrier momentums in zero potential layers, that characterize the QP state and called also average momentums of the QP, are not available. In the zero-doping ground state at $T<T^*$ electrons fill permitted QP states with the minimal (among unfilled yet states) energy until the occupied area reaches a half of the FBZ area $b^2/2$. As in the areas shown with grey in Fig. 8(a) the QP states are forbidden in presence of CO potential (more precisely, the angle γ corresponding to PG closing and shown in Fig. 8(a) as well as $φ_{clos}$ slightly depends on the QP energy as is seen from Fig.5(b,c), but here we neglect this dependence), the FS corresponding to zero doping is shifted to higher QP energies at low temperatures $T<T^*$ in comparison with its position at high temperature $T>T^*$. One can note similar behavior of the FS comparing experimental antinodal ARPES spectra obtained at low and high temperatures [6]. The low- and high-temperature zero-doping FSs are shown with dashed blue and red lines, respectively, in Fig. 8(a). Since according to low-temperature experimental data the dispersion should be the most flat near the FS [4,38,54], in the model dispersion we place the most flat region near the low-temperature zero-doping FS. Blue and red solid lines in Fig. 8(a) represent the FSs corresponding to p=0.15 at $T<T^*$ and $T>T^*$, respectively.

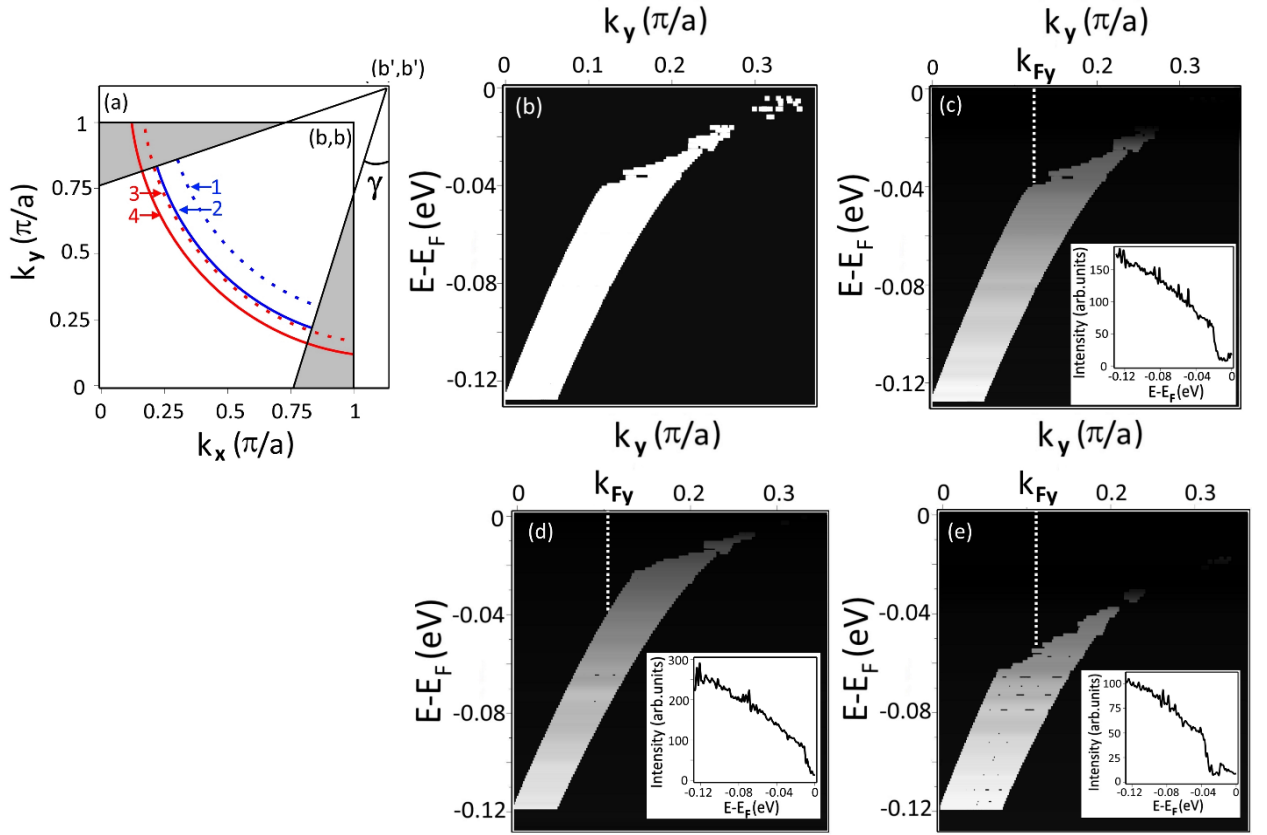

FIG. 8. (a) Areas of the FBZ unavailable for QPs at low temperatures $T<T^*$ are shown with grey; (b',b') is a center of arcs representing constant energy surfaces (curves); dashed lines 1 and 3 depict zero-doping FS at $T<T^*$ and at $T>T^*$, respectively, solid lines 2 and 4 present FS for the doping $p=0.15$ at $T<T^*$ and at $T>T^*$, respectively; (b) white color shows areas where spectral weight in antinodal ARPES spectrum differs from zero, so far without differentiation in the intensity; (c), (d) and (e) calculated antinodal ARPES spectra for $U_0$=0.05, 0,04 and 0.065 eV, respectively, dashed line shows the PG width determined as midpoint shift at $k_F$, where $k_F$ is Fermi momentum at $T>T^*$, inset represents relative intensity (calculated as inverse area of the momentum-space trajectory for QP with the energy $E$) as function of binding energy,

Constructing an ARPES spectrum in the form of image plot (for comparison with experimental image plot [6]) we first take a thin stripe which is intersection of the momentum-space trajectory of a QP with the energy $E$ and the FBZ boundary and put it in the corresponding place of the plane ($k_y$, $E$-$E_F$). In such a way we obtain areas with non-zero spectral weight demonstrated by Fig. 8(b) where the intensity variations are not taken into account so far. Photoemission intensity for given energy $E$ and momentum $k_y$ (the probability of the QP with the energy $E$ to escape with fixed $y$-projection of the momentum $k_y$) obviously depends on the area of the momentum-space trajectory. Due to the flatness of the cuprates-like dispersion near the FS the trajectory area depends strongly on the QP energy as it is seen in Fig. 7(b). The intensity calculated as inverse area of the momentum-space trajectory for QP with the energy E is presented by inset in Fig. 8 (c).

Fig. 8(c) shows the resulting antinodal ARPES spectrum (image plot) with taking into account the intensity variation at various binding energy. The spectrum clearly demonstrates presence of the PG, the PG width determined as midpoint shift at the Fermi momentum $k_{Fy}$ (momentum corresponding to Fermi crossing at high temperature $T>T^*$, shown in Fig. 8(c)) is slightly lower than the CO potential amplitude $U_0$ ($U_0$=0.05 eV in the Fig. 8(c)). Spectrum demonstrates giant broadening in energy, as it takes place in the experiments [6,7]. Long continuation of the spectral weight at $k>k_F$ demonstrated by the calculated spectrum in Fig. 8(c) is also observed in experiments [6,7]. Often this experimental peculiarity is referenced as $k_F$ misalignment or particle-hole symmetry breaking [6-8,15,16]. FS crossing is not observed in the calculated spectrum (as in experimental spectra) due to large increase of the trajectory area near FS (caused by flatness of the band near FS) and corresponding large decrease in the intensity.

Experimental STM spectra show that there is some dispersion of the PG width in one and the same cuprate crystal, likely due inhomogeneity caused by presence of dopant ions [13,14]. Therefore, we calculate antinodal ARPES spectra for $U_0$=0.065eV and 0.04 eV, they are presented by Fig.8(d) and (e), respectively. If to take into account the inhomogeneity, ARPES experiment will see the superposition of the spectra shown in Fig.8(c,d,e) and spectra with $U_0$ values in-between (with corresponding weights, that are doping dependent). This will result in gradual changing the spectral weight along $k_y$ axis and additional broadening.

### 3.4. Display of DWPs in a system with cuprates-like dispersion in STM spectra

Finally let us briefly discuss manifestation of the described above QPs in scanning tunneling microscopy (STM) which is the other basic experimental method of the PG study [10-14]. Considering the differential conductance mode, it is convenient to use the fact that differential conductance is proportional to the carrier local density of states (LDOS) [12,57]. Obviously, in the present approach LDOS gradually increases with the bias $V$ up to its value $U_0/e$, since near-antinodal states that are unavailable at zero bias (due to presence of negative half-wave of the potential energy) gradually come to play through lowering the closing angle $\varphi_{clos}$ at increasing bias. With the increase of bias, near-antinodal states become available as carriers acquire lacking energy (which varies from 0 up to $U_0$ for states with different "angular distance" to the antinode) from the applied bias field.

Thus, compensating negative potential energy in CO potential by the carrier energy $eV$ in the applied field results in increasing density of available states. The same is correct for the hole states at bias equal in the absolute value but opposite in the sign. Such increase of differential conductance (representing density of available states) with increasing bias up to the PG width value is observed in experiments on cuprates [10-14]. Experiments demonstrate also spatial variation of the PG width [12-14,58] ordinarily related with inhomogeneities caused by dopant ions. The present approach allows calculating the potential amplitude with taking the potential of dopant ions into account, that may be a subject for future study.

At considering the topographic mode in the frames of the present approach the bias is superimposed on negative or positive CO potential in different regions of the surface. This results in variation of the tip height necessary for constant tunneling current, thus visualizing the charge distribution, that is apparently seen in experiments. Obviously, the absence of near-antinodal QPs due to impact of CO potential displays itself as a PG also in the specific heat, thermal conductivity, Knight shift and other properties depending on the carrier density of states near FS.

### 3.5. Calculated doping dependence of the PG width and PG onset temperature

The present approach enables one to calculate doping dependence of the PG width and PG onset temperature $T^*$. As is demonstrated above, in the approach the PG width measured with STM is equal to the CO potential amplitude $U_0$, the PG width in antinodal ARPES spectrum is slightly lower than $U_0$. Doping dependence of the CO potential amplitude is caused by doping dependence of the CO wave vector projections $K_{COx,y}$ [44] entering Eqs. (1,2). Using experimental $K_{COx,y}(p)$ dependence [44] in Bi2212 and Eqs. (1,2) we calculate the CO potential amplitude $U_0$ as function of doping. It is presented in Fig. 9 (a) together with PG width values measured in Bi2212 experimentally with STM [11]. Comparison shows agreement of the calculated doping dependence of the PG width with the experimental one.

Fig. 9(b) represents doping dependence of the PG onset temperature $T^*$ calculated in the present approach as temperature corresponding to thermal decay of 95% of bipolarons responsible for the CO potential. To obtain the temperature of the bipolarons thermal decay we use distribution function of carriers in systems with strong long-range EPI [59] (describing possible coexistence of autolocalized and delocalized carriers) and bipolaron energy as function of doping. The latter was calculated with variation method generalized to a system of coexisting bipolaron liquid and delocalized carriers [41]. To obtain bipolaron density according to the distribution function [59] the maximum group velocity u of phonons strongly coupled with charge carriers is needed. Its value ($u$=1000 ms$^{-1}$) was estimated from the phonon dispersion in the vicinity of the CO wave vector in cuprates [24]. As is seen from Fig. 9(b) the temperature $T^*$ of the PG onset/disappearance calculated in the present approach is in good agreement with that observed in LSCO and Nd/Eu-LSCO [17].

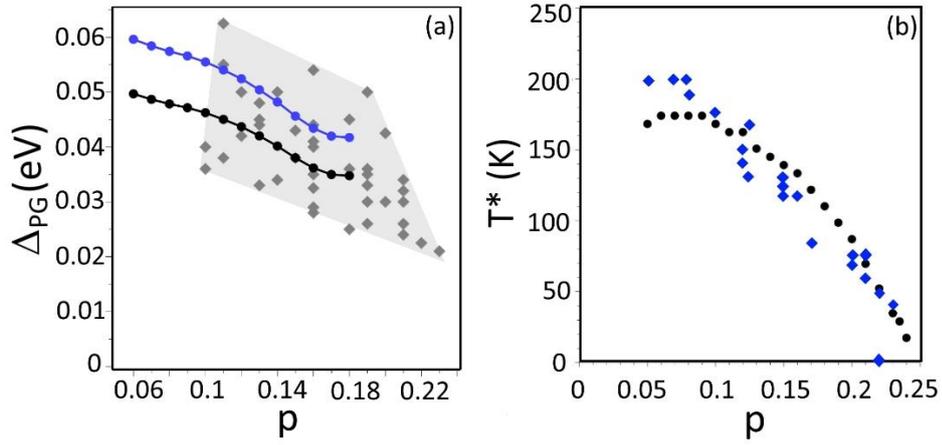

FIG. 9. (a) Blue and black circles represent doping dependence of the CO potential amplitude $U_0(p)$ (equal in the present approach to the PG width measured with STM methods) calculated according to (1,2) with using experimental $K_{CO}(p)$ dependence [44] for Bi2212, at $r_0$=200Å, $\varepsilon_0$=50 and 60, respectively; grey rhombuses show experimental data obtained with STM on Bi2212 [11]; (b) doping dependence of the PG onset temperature $T^*$ calculated in the present approach according to [41,59] as temperature corresponding to thermal decay of 95% of bipolarons at dielectric constant $\varepsilon_0$=30, inverse effective dielectric permittivity $1/\varepsilon^*$=0.27 and phonon group velocity $u$=1000 ms$^{-1}$ (black circles) and experimental PG onset temperature observed in LSCO and Nd/Eu-LSCO [17] (blue rhombuses).

### 3.6. Resulting ground normal state and FS reconstruction

Let us summarize features of the ground normal state of system with cuprates-like dispersion and strong Frohlich EPI in different parts of doping-temperature phase diagram. Ordinarily several characteristic temperatures and doping levels are introduced to describe a variety of phases in hole-doped cuprates [1-3,17,18,44,60]. Let us discuss the difference between some of these phases in the present approach. At temperature $T<T^*$ the system demonstrates PG behavior, that displays itself in ARPES and STM spectra, in other properties depending on the carrier LDOS. This behavior takes place in the present approach until thermal decay of the most part (here about 95%) of hole bipolarons. Doping dependence of $T^*$ stems from doping dependence of the bipolaron binding energy [41].

The relation between PG and CO in cuprates is under discussion for long time. In the present approach the potential generated by bipolarons (denoted also as CO potential) is one of sources of the PG together with dispersion topology. The reason of experimentally observed relation between the temperatures $T_{CO}$ (the temperature of CO decay/onset) and $T^*$, $T_{CO}<T^*$, in the considered system is quite clear. Resonant elastic X-ray scattering (REXS) is one of the main tools to study CO in cuprates [52,44]. Calculation of the REXS spectrum from CO formed by bipolarons [41] has shown that $n$ times decrease in the number of bipolarons results in approximately $n^2$ times decrease in the REXS intensity at $K_{CO}$. Thus, for CO to be observed the much larger (than 5%) portion of bipolarons from their number at $T$=0K should be present. According to distribution function for systems where autolocalized and delocalized carriers can coexist [59] the temperatures of decay of a half bipolarons and 95% of bipolarons differ significantly, the difference depends on doping, bipolaron energy and group velocity of phonons strongly coupled with charge carriers. The relation between CO and PG in the model under study displays itself also in relation of doping dependences of the CO wave vector $K_{CO}(p)$ and PG width $\Delta_{PG}(p)$: calculation according to (1,2) with using experimentally measured $K_{CO}(p)$ in Bi2212 [44] yields $\Delta_{PG}(p)$ that agrees with experiments on the same Bi2212 system [17].

The developed idea about ground normal state where only autolocalized holes (until rather high doping is reached) are present and electron near-antinodal DWPs are absent sheds light on the reason of the FS reconstruction from small electron pocket into large hole-like FS (enclosing the area proportional to $p+1$) observed at some doping $p=p^*$ in measurements of the frequency of quantum oscillations in magnetic field [18,3]. At low doping delocalized holes are absent whereas electron QPs have closed trajectories in the magnetic field, restricted by PG presence as was revealed in Ref. [18]. At $p<p^*$ only the electron FS with rather small area due to absence of near-antinodal QPs will occur as was suggested and confirmed with measurements of quantum oscillation frequency by Harrison and Sebastian [18]. (The source of sign change of the DWP average wave vector projections in the present approach is scattering on the CO potential.) Upon increase of doping the bipolaron binding energy decreases and at sufficiently high doping $p=p^*$ becomes close to zero so that thermal decay of bipolarons occurs at very low temperatures as is illustrated by Fig.9(b). Thus, at $p>p^*$ delocalized holes appear, they have closed trajectories in magnetic field, the area enclosed by the FS is proportional to $p+1$, as is observed in experiments [2,3,18]. Approximately at the same $p=p^*$ PG disappears due to bipolarons decay. Ref. [17] introduces the other characteristic doping level $p_{FS}$ corresponding to Lifshits transition when the FS becomes electron-like (thus stressing importance of topology of carrier constant energy surfaces). In the present approach Lifshits transition also results in PG disappearance as is illustrated by Fig.4 (c) (if $p_{FS}<p^*$).

## 4. CONCLUSION

In summary, we demonstrate that potential created by autolocalized carriers (called also charge ordering potential as they are arranged in such a way) reconstructs Bloch stationary states into DWPs with different momentums in areas with different potential. Specific topology of 2D carrier dispersion in hole-doped cuprates results in absence of DWPs with average momentums near antinode. Nevertheless, photoelectrons with antinodal momentums are observed in such systems. They stem from DWPs with average momentums rather far from antinode when the photoelectron escapes from areas with minimal (negative) potential energy in the charge ordering potential. Therefore, the energy of such photoelectrons is by approximately the amplitude of the additional potential energy lower than that dictated by dispersion for their momentums. As a result, spectral weight in ARPES spectra is shifted to higher (in the absolute value) binding energy and greatly broadened, just as in experiments on cuprates. Besides, the spectral weight appears at momentums higher than the Fermi momentum but with energies well below the Fermi energy, that is also characteristic for near-antinodal ARPES spectra of cupartes. Fermi crossing nevertheless is not observed due to great decrease in intensity near Fermi surface caused by flat band in this region.

Apart of ARPES spectra, absence of near-antinodal QPs displays itself in STM spectra and other properties depending on the carrier density of states. Calculated doping dependence of the PG width and PG onset temperature shows agreement with experimental studies of PG in cuprates. The study changes conventional idea about ground normal state of cuprates. In the suggested approach doped holes are autolocalized at doping $p$ lower than some value $p^*$ whereas delocalized QPs represent DWPs and cannot exist with average momentums near antinodes. This explains Fermi surface reconstruction from small electron pocket into large hole-like Fermi surface observed in cuprates at $p=p^*$ with quantum oscillation measurements. The suggested relation between the charge ordering and PG is in harmony with sequence of their onset temperatures.

In future, the developed approach to ground normal state of hole-doped cuprates and understanding changes in their density of states may help to shed light on complex Hall constant behavior as function of doping and temperature, spatial variation of the pseudogap width observed in STM spectra, well-known enigma of peak-dip-hump structure in ARPES and STM spectra and other peculiarities observed in cuprates. From the other hand, the results obtained open a possibility of creating systems with artificial pseudogap and switchable density of states on the base of highly polarizable layered structures with 2D conductivity and hole-like dispersion.

**Acknowledgments**

We are grateful to A. S. Mishchenko, D. Emin, S. G. Ovchinnikov and S B. Rochal for valuable communication and useful comments. The work was supported by Southern Federal University.